# Determination of $e^+e^-$ Beam Polarization for Optimal Gauge Boson Discovery Limits

Ali A. Bagneid[a]

*Department of Physics, Umm Al-Qura University, Makkah, Saudi Arabia*

**Abstract**

Extra neutral gauge bosons suggested by models beyond the standard model can indirectly show up in $e^+e^-$ collisions through off-resonance deviations of various physical observables from the corresponding standard model values. We considered leptonic observables and studied the dependence of the deviations on the polarizations of the positron and electron beams. We showed that, for a given model, the magnitude of the deviation of a given observable can attain its maximum value if the polarizations of the positron and electron beams are properly chosen. We determined, for a given model, a single set of beam polarization so that if this set is employed in measuring all considered observables, it produces the highest extra gauge boson discovery limits.

____________________

[a] E-mail: bagneid@yahoo.com

At the present time, the standard model (SM) provides a remarkably successful framework for describing the observed elementary particle interactions at all the energies thus far probed. However, there are questions that cannot be answered satisfactorily within the framework of the standard model. For example, why is the gauge structure a product of three gauge groups rather than one? Why are there three generations of fermions? etc. Several models have been proposed to address these issues. Such theories may have large symmetry G, which contains the standard model gauge symmetry, $G_{SM} = SU(3)_c \otimes SU(2)_L \otimes U(1)_Y$, as a subgroup. Examples are, grand unified theories, super-symmetric models, superstring models, etc. Most of these theories predict the existence of extra neutral gauge bosons. So far, null search results are obtained for extra gauge bosons at all available energies. Experimentally, we try to look for new phenomena that result from the existence of these particles. One possible direct way is to look for on-resonance signals for the new gauge bosons. In this context, the large hadron collider (LHC) at CERN is expected to be capable of probing gauge bosons of masses in the few TeV range. If the gauge boson mass is beyond this range, it will be necessary to push the collision energy higher by building new hadron colliders. Another option is to seek indirect means in searching for the new particles. Here we try to look for possible effects of these particles that would cause small off-resonance deviations of various physical observables from the SM values. The indirect identification may be difficult in hadronic colliders because of limited statistics. This option is more appropriate for electron-positron colliders, where high precision measurements of any extra gauge bosons for their couplings and interactions with other particles, are expected. Moreover, it allows us to probe gauge boson masses much higher than those allowed directly at hadronic colliders. We will be interested here in this indirect option. It is desirable to push the discovery limits of the extra gauge bosons by enlarging the stated deviations through proper adjustment of relevant parameters. We will consider in this work possible off-resonance deviations of leptonic observables in high-energy electron positron collisions caused by the existence of extra neutral gauge bosons suggested by models beyond the SM. We intend to show that the magnitudes of these deviations can attain their maximum values by properly adjusting characteristics of the colliding beams.

Any deviations of observables from the SM values are caused jointly by the model parameters and by parameters belonging to the colliding beams. The model parameters are the mass $M^\alpha$ of the extra neutral gauge boson $Z_2 \equiv Z_\alpha$ belonging to model $\alpha$, its vector coupling, $v_2^f$, and axial-vector couplings, $a_2^f$, to fermion $f$, and its mixing angle, $\varphi$, with the SM gauge boson, $Z_1 \equiv Z_{SM}$. The colliding beam parameters are the collision energy, $\sqrt{s}$, the degrees of longitudinal polarization, $\lambda_+$ and $\lambda_-$, of respectively the positron and electron beams, and their luminosity, $L$. For a given model $\alpha$, $M^\alpha$ is unknown, $v_2^f$ and $a_2^f$ are fixed, and $\varphi \approx 0$ [1]. On the other hand, in principle, there are no restrictions on the colliding beam parameters. For sufficiently large $M^\alpha$, $M^\alpha \gg \sqrt{s}$, the magnitudes of the deviations are expected to increase with $\sqrt{s}$. The deviations are also expected to increase with $L$ because of the decrease in the experimental errors associated with the observables. However, an $e^+e^-$ collision experiment is usually designed to operate within a limited range of collision energies and luminosities. Raising the collision energy further requires building new colliders while increasing the luminosity requires new techniques. Thus, the parameters that we are left with are the polarizations, $\lambda_+$ and $\lambda_-$, of the colliding beams. In what follows we will first attempt to determine for each observable as predicted by a given model, a set of beam polarization that maximizes the magnitude of its deviation from the SM value. Let us take, as a case study, the neutral gauge bosons $Z_{LR}$ suggested by left-right symmetric models (*LR*) [2] and $Z_\alpha$, $\alpha = \chi, \psi, \eta, I'$, suggested by grand unified theories based on $E_6$ and *SO*(10) groups (including superstring models) [3]. We will consider here the proposed international linear collider (ILC) [4] which is expected to be the next $e^+e^-$ linear collider. The ILC is designed to collide electrons and positrons with collision energies in the range $0.5-1$ TeV. The polarizations of the electron and positron beams at the ILC are expected to reach a high degree of at least 80% and 60% polarization, respectively. We will consider the following observables: the total cross section, $\sigma(\mu) \equiv \sigma(e^+e^- \to \mu^+\mu^-)$, the forward-backward asymmetry, $A_{FB}(\mu) \equiv A_{FB}(e^+e^- \to \mu^+\mu^-)$ and the left-right asymmetry, $A_{LR}(\mu) \equiv A_{LR}(e^+e^- \to \mu^+\mu^-)$. Three sets of beam polarizations will be of interest here among others. The first set, denoted by $U$, is a combination of the set of unpolarized beams, $(\lambda_+, \lambda_-) = (0\%, 0\%)$, to be used in measuring the unpolarized total cross section and forward-backward asymmetry, and the set, $(\lambda_+, \lambda_-) = (0\%, -80\%)$, to be used in measuring the left-right asymmetry. The second set, denoted by $P$, is a single set in which the polarizations of the positron and electron beams are given by: $(\lambda_+, \lambda_-) = (60\%, -80\%)$. The third set, denoted by $R$, is also a single set in which the signs of the polarizations of the positron and electron beams of the second set are reversed where, $(\lambda_+, \lambda_-) = (-60\%, 80\%)$.

The general expression for the differential cross section [5] as predicted by model $\alpha$ for the process $e^+e^- \to \mu^+\mu^-$, when longitudinally polarized beams are employed and the helicities of the final states measured is given by:

$$\frac{d\sigma^\alpha}{d\Omega} = \frac{\alpha^2}{4s}\left[A(1+\cos^2\vartheta) + 2B\cos\vartheta\right], \qquad (1)$$

where $\vartheta$ is the angle between the incident electron and the outgoing muon and,

$$A = \lambda_1 (h_1 F_1^\alpha + h_2 F_4^\alpha) + \lambda_2 (h_1 F_3^\alpha + h_2 F_2^\alpha), \qquad (2)$$

$$B = \lambda_1 (h_1 F_2^\alpha + h_2 F_3^\alpha) + \lambda_2 (h_1 F_4^\alpha + h_2 F_1^\alpha). \qquad (3)$$

Here $\lambda_1 = 1 - \lambda_+ \lambda_-$, $\lambda_2 = \lambda_+ - \lambda_-$, $h_1 = (1 - h_+ h_-)/4$ and $h_2 = (h_+ - h_-)/4$, where $h_+$ and $h_-$ denote the helicities of the final $\mu^+$ and $\mu^-$ leptons, respectively. In this work we will be concerned only with the polarizations of the initial state leptons. With $v_j^e = v_j^\mu \equiv v_j$ and $a_j^e = a_j^\mu \equiv a_j$, $j = 1,2$, where the couplings $v_1$ and $a_1$ belong to the SM while $v_2$ and $a_2$ belong to model $\alpha$, the functions $F_n^\alpha$, $n = 1,2,3,4$ read:

$$F_1^\alpha = 1 + 2 \sum_{j=1}^{2} (v_j)^2 \chi_j + \sum_{j,k=1}^{2} (\chi_j \chi_k + \eta_j \eta_k)(v_j v_k + a_j a_k)^2, \qquad (4)$$

$$F_2^\alpha = 2 \sum_{j=1}^{2} (a_j)^2 \chi_j + \sum_{j,k=1}^{2} (\chi_j \chi_k + \eta_j \eta_k)(v_j a_k + a_j v_k)^2, \qquad (5)$$

$$F_3^\alpha = F_4^\alpha = 2 \sum_{j=1}^{2} a_j v_j \chi_j + 2 \sum_{j,k=1}^{2} (\chi_j \chi_k + \eta_j \eta_k) v_j a_k (v_j v_k + a_j a_k). \qquad (6)$$

The functions $\chi_j$ and $\eta_j$, $j = 1,2$ are given by:

$$\chi_j = \frac{s(s - M_j^2)}{x_w (1 - x_w)\left[(s - M_j^2)^2 + M_j^2 \Gamma_j^2\right]}, \qquad (7)$$

$$\eta_j = \frac{-s M_j \Gamma_j}{x_w (1 - x_w)\left[(s - M_j^2)^2 + M_j^2 \Gamma_j^2\right]}. \qquad (8)$$

Here $x_w \equiv \sin^2 \vartheta_W$ whereas $M_j$ and $\Gamma_j$ are the mass and total width [6] of the gauge boson $Z_j$, respectively. Let us examine the dependence of $d\sigma^\alpha/d\Omega$ on the various parameters. The differential cross section depends on $\lambda_+$ and $\lambda_-$ through the combinations, $\lambda_1 = 1 - \lambda_+ \lambda_-$ and $\lambda_2 = \lambda_+ - \lambda_-$. $d\sigma^\alpha/d\Omega$ is independent of the luminosity, $L$, where as we stated above $L$ can only affect the experimental errors associated with the observables. The rest of the parameters are represented in the functions $F_n^\alpha$, $n = 1,2,3,4$ which we here parameterize as follows:

$$F_n^\alpha = \delta_{1,n} + \sum_{j=1}^{2} K_j^n + \sum_{j,k=1}^{2} K_{jk}^n, \qquad (9)$$

where the terms $K_j^n$ and $K_{jk}^n$ are given by:

$$K_j^n = C_j^n \chi_j, \qquad (10)$$

$$K_{jk}^n = C_{jk}^n (\chi_j \chi_k + \eta_j \eta_k), \qquad (11)$$

and the functions $C_j^n$ and $C_{jk}^n$ can easily be read from Eqs. (4)-(6). The model parameters affect the magnitude and sign of the deviation of a given observable from the corresponding SM value through the functions, $F'^\alpha_n$, $n = 1,2,3,4$ [7], where:

$$F'^\alpha_n = K_2^n + K_{12}^n + K_{21}^n + K_{22}^n. \qquad (12)$$

In order to obtain discovery limits for an extra gauge boson $Z_\alpha$, we construct the $(\chi^2)^\alpha$ figure of merit:

$$(\chi^2)^\alpha = \sum_{i=1}^{3} \left( \frac{\Delta O_i^\alpha}{\delta O_i^{SM}} \right)^2 \equiv \sum_{i=1}^{3} (\chi^2)_i^\alpha, \qquad (13)$$

where,

$$\Delta O_i^\alpha = O_i^\alpha - O_i^{SM}. \qquad (14)$$

Here $O_i^\alpha$ represents the value of the observable $O_i$ as predicted by model $\alpha$, where $O_i^\alpha = \sigma^\alpha(\mu), A_{FB}^\alpha(\mu), A_{LR}^\alpha(\mu)$, for $i = 1,2,3$, respectively. $O_i^{SM}$, $i = 1,2,3$, are the corresponding SM predictions and $\delta O_i^{SM}$ is the experimental error associated with $O_i^{SM}$. All parameters mentioned above are contained in the components $(\chi^2)_i^\alpha$, $i = 1,2,3$. In particular, the mass, $M^\alpha$, of the extra gauge boson $Z_\alpha$, is represented in $O_i^\alpha$ via the functions $F_n'^\alpha$, $n = 1,2,3,4$. We are concerned here with very small deviations that result from very large gauge boson masses. We note that, as $M^\alpha$ becomes sufficiently larger than $\sqrt{s}$, the functions $F_n'^\alpha$, $n = 1,2,3,4$ become inversely proportional to $(M^\alpha)^2$. Discovery limits for an extra gauge boson, $Z_\alpha$, are obtained by varying $M^\alpha$ in $(\chi^2)^\alpha$ and comparing the predictions of the observables, assuming the presence of the extra gauge boson, to the predictions of the SM [8]. As we stated above, we wish to find for each observable as predicted by a given model, a set of beam polarization that maximizes the magnitude of its deviation from the corresponding SM value. These sets should also maximize the discovery limits of the extra gauge boson. We allowed the polarizations $\lambda_+$ and $\lambda_-$ to vary within their allowed ranges while keeping the other parameters, including $M^\alpha$, fixed and calculated for each observable $O_i^\alpha$, $i = 1,2,3$, as predicted by model $\alpha$, a set of beam polarization, $(\lambda_+, \lambda_-)_{i,Max}^\alpha$, that maximizes the corresponding component $(\chi^2)_i^\alpha$ of $(\chi^2)^\alpha$. The algebraic structure of the components $(\chi^2)_i^\alpha$ tells us that if these sets are then used as input and we allowed $M^\alpha$ to vary while keeping all other parameters fixed, the rate at which the components $(\chi^2)_i^\alpha$ decrease with $M^\alpha$ will be at its minimum. The discovery limits for $Z_\alpha$ obtained using the sets $(\lambda_+, \lambda_-)_{i,Max}^\alpha$, $i = 1,2,3$, will thus be maximal. In other words, the sets $(\lambda_+, \lambda_-)_{i,Max}^\alpha$, $i = 1,2,3$, are expected to be the most efficient if used in measuring the corresponding observables, $O_i^\alpha$, $i = 1,2,3$, in an experiment designed to search for an extra neutral gauge boson, $Z_\alpha$, belonging to model $\alpha$.

In table 1 we display for each model, the corresponding sets, $(\lambda_+, \lambda_-)^{\alpha}_{i,Max}$, $i = 1,2,3$. We similarly calculated for each model the single set, $(\lambda_+, \lambda_-)^{\alpha}_{max}$, of beam polarization that maximizes $(\chi^2)^{\alpha}$ and obtained $(\chi^2)^{\alpha}_{max}$, where:

$$(\chi^2)^{\alpha}_{max} = \left(\sum_i (\chi^2)^{\alpha}_i\right)_{max}. \tag{15}$$

The results for $(\lambda_+, \lambda_-)^{\alpha}_{max}$ are also displayed in table 1. In our calculations we used $\sqrt{s} = 0.5$ TeV, $L = 0.5$ ab$^{-1}$, $M^{\alpha} = 2$ TeV and considered only statistical errors. In order to explain our results, we will also need to obtain the sets of beam polarizations, $(\lambda_+, \lambda_-)^{SM}_{j,Max}$ and $(\lambda_+, \lambda_-)^{SM}_{j,Min}$, where $j = \sigma, FB, LR$, that respectively maximize and minimize the magnitude of each of the SM observables $\sigma^{SM}(\mu)$, $A^{SM}_{FB}(\mu)$ and $A^{SM}_{LR}(\mu)$. We found that $(\lambda_+, \lambda_-)^{SM}_{j,Max} = P$ for $j = \sigma, FB$ and $(\lambda_+, \lambda_-)^{SM}_{LR,Max} = P, R$, where the magnitude of the left-right asymmetry is maximized by any of the sets, $P$ or $R$. On the other hand, $(\lambda_+, \lambda_-)^{SM}_{\sigma,Min} = (60\%, 80\%)$, $(\lambda_+, \lambda_-)_{FB,Min} = R$, and the condition $\lambda_+ = \lambda_-$ minimizes the magnitude of the left-right asymmetry. Here also we used $\sqrt{s} = 0.5$ TeV, $L = 0.5$ ab$^{-1}$ and considered only the statistical errors.

We will attempt to explain the results of table 1 by investigating the individual components, $(\chi^2)^{\alpha}_i$, $i = 1,2,3$, of $(\chi^2)^{\alpha}$. These components depend on the quantities, $\Delta O^{\alpha}_i$ and $\delta O^{SM}_i$. It is possible that the set of beam polarization that maximizes $\Delta O^{\alpha}_i$ be different from the set that minimizes $\delta O^{SM}_i$ and that only one of these two quantities determines $(\lambda_+, \lambda_-)^{\alpha}_{i,Max}$. This appears to be the case for $(\chi^2)^{\alpha}_1$. The set of beam polarization that minimizes $\delta O^{SM}_1$ is the set that minimizes $\sigma^{SM}$, namely, $(\lambda_+, \lambda_-)^{SM}_{\sigma,Min} = (60\%, 80\%)$. Table 1 shows that the sets, $(\lambda_+, \lambda_-)^{\alpha}_{1,Max}$ $\alpha = \chi, \psi, \eta, LR, I'$, are all different from $(\lambda_+, \lambda_-)^{SM}_{\sigma,Min}$. Now, knowing that,

$$(\Delta O_1^{\alpha})^2 \propto \left(F_1'^{\alpha}\right)^2 \left[ (\lambda_1)^2 + (\lambda_2)^2 \left(\frac{F_3'^{\alpha}}{F_1'^{\alpha}}\right)^2 + 2\lambda_1 \lambda_2 \left(\frac{F_3'^{\alpha}}{F_1'^{\alpha}}\right) \right], \quad (17)$$

and that both polarization sets $P$ and $R$ maximize the magnitudes of $\lambda_1$ and $\lambda_2$, the values of $\lambda_+$ and $\lambda_-$ that maximize the magnitude of the deviation $\Delta O_1^{\alpha}$ are those values that maximize the quantity $\lambda_1 \lambda_2 (F_3'^{\alpha}/F_1'^{\alpha})$. The set $P$ ($R$) is therefore the appropriate choice for a positive (negative) $(F_3'^{\alpha}/F_1'^{\alpha})$. In figure 1 we show the quantities $(F_3'^{\alpha}/F_1'^{\alpha})$, as functions of $M^{\alpha}$ for the considered models. According to Fig. 1 the appropriate polarization sets are $P$ for models $\chi$, $\psi$ and $I'$ and $R$ for models $\eta$ and $LR$, in agreement with the results given in table 1. Thus, it turns out that the deviations, $\Delta O_1^{\alpha}$, do indeed determine $(\lambda_+, \lambda_-)^{\alpha}_{1,Max}$ for our models.

Although the set $P$ of beam polarization is preferred by $\delta O_2^{SM}$, where it simultaneously maximizes $\sigma^{SM}$ and the magnitude of $A_{FB}^{SM}$, the final decision on the most appropriate set of beam polarization, $(\lambda_+, \lambda_-)^{\alpha}_{2,Max}$, is to be taken jointly with $\Delta O_2^{\alpha}$. The set of beam polarization that maximizes $\Delta O_2^{\alpha}$ depends on the magnitudes and signs of the functions $F_n'^{\alpha}$, $n = 1,2,3,4$, $\lambda_1$ and $\lambda_2$. However, this dependence is not as transparent as in the case of the cross section. The results listed in table 1 show that $(\lambda_+, \lambda_-)^{\alpha}_{2,Max}$ equals $P$ for models $\eta$ and $I'$ and equals $R$ for models $\chi$, $\psi$ and $LR$.

As for $(\chi^2)^{\alpha}_3$, the set of beam polarization that minimizes $\delta O_3^{SM}$ is that set that simultaneously maximizes the absolute value of the effective polarization, $|P_{eff}|$, where $P_{eff} = \lambda_2/\lambda_1$, and $\sigma^{SM}$. On the other hand, the polarization set that maximizes $\Delta O_3^\alpha$ is the set that maximizes $|P_{eff}|$ where :

$$(\Delta O_3^\alpha)^2 = (P_{eff})^2 \left( \frac{F_3^{SM} + F_3^{'\alpha}}{F_1^{SM} + F_1^{'\alpha}} - \frac{F_3^{SM}}{F_1^{SM}} \right)^2, \qquad (18)$$

and $F_n^{SM} = F_n^\alpha - F_n^{'\alpha}$, $n = 1,2,3,4$. There is only one candidate that satisfies these conditions, namely, the polarization set $P$. Obviously the set $(\lambda_+, \lambda_-)_{3,Max}^\alpha$ is independent of the model parameters.

We thus calculated for each model the sets $(\lambda_+, \lambda_-)_{i,Max}^\alpha$, $i = 1,2,3$, and determined the maximum values, $(\chi^2)_{i,Max}^\alpha$ of the individual components of $(\chi^2)^\alpha$, and obtained $(\chi^2)_{Max}^\alpha$ where:

$$(\chi^2)_{Max}^\alpha = \sum_{i=1}^{3} (\chi^2)_{i,Max}^\alpha. \qquad (19)$$

We have also used the single sets of beam polarization, $U$, $P$ and $R$, as input and calculated for the models the corresponding values of $(\chi^2)_U^\alpha$, $(\chi^2)_P^\alpha$ and $(\chi^2)_R^\alpha$. In figure 2 we present the stack column graphs of $(\chi^2)_p^\alpha$, $p = U, P, R, M$, for all models, where we used the symbol $M$ to refer to the sets: $(\lambda_+, \lambda_-)_{i,Max}^\alpha$, $i = 1,2,3$. In our calculations we used $\sqrt{s} = 0.5$ TeV, $L = 0.5$ ab$^{-1}$, $M^\alpha = 2$ TeV and considered only statistical errors. Fig. 2 not only compares the values of the different $\chi^2$'s, but shows also the relative importance of each of their components. The figure shows that for a given model $\alpha$, $(\chi^2)_{Max}^\alpha$ has the highest among $(\chi^2)_p^\alpha$, $p = U, P, R, M$, as expected. Since the single sets, $(\lambda_+, \lambda_-)_{max}^\alpha$, turned out to be either the set $P$ or $R$, the stack columns graphs for $(\chi^2)_{max}^\alpha$ are indeed represented in Fig.

2, where we underlined, for each model, the polarization symbol that represents $(\lambda_+, \lambda_-)^{\alpha}_{\max}$.

In what follows we will examine our results by calculating discovery limits for the considered extra neutral gauge bosons. In our calculations we will employ three sets of colliding beam parameters, $(\sqrt{s}, L)$, and three sets of error scenarios. In the first set of colliding beam parameters, which we will refer to as set $A$, we chose, $\sqrt{s} = 0.5$ TeV and $L = 0.5$ ab$^{-1}$, while in set $B$ $\sqrt{s} = 0.5$ TeV and $L = 1$ ab$^{-1}$ and in set $C$ $\sqrt{s} = 1$ TeV and $L = 1$ ab$^{-1}$. Since the discovery limits depend on deviations from the SM predictions, they are sensitive to both statistical and systematic errors. We calculated the total error by combining in quadrature the statistical error and the systematic error. We assumed equal relative precision, $x \equiv \delta\lambda_+/\lambda_+ = \delta\lambda_-/\lambda_- = 0.5\%$ of the two beam polarizations and calculated the error in the effective polarization under the assumption that the errors are completely independent and added in quadratures [9]. We will consider three error scenarios. In the first scenario, which we refer to as scenario $a$, we consider only the statistical error. In scenarios $b$ and $c$, we take $\delta\lambda_\pm/\lambda_\pm = 0.5\%$, a 0.1% error for the measurement of each observable and, respectively, a 0.5%(0.2%) error in the cross section measurements due to, for example, luminosity uncertainties, and a corresponding 0.25%(0.1%) one in asymmetries where errors partially cancel.

Before using our calculated sets of beam polarizations, it is necessary to check their dependence on the chosen sets of collider beam parameters, error scenarios and, when applicable, $M^\alpha$. The sets of beam polarizations, $(\lambda_+, \lambda_-)^{SM}_{j,Max}$ and $(\lambda_+, \lambda_-)^{SM}_{j,Min}$, $j = \sigma, FB, LR$ given above were calculated using the set $(A, a)$ of colliding beam parameters $A$ and error scenario $a$. We repeated the calculations but for all other different sets, $(A-C, a-c)$, of the chosen colliding beam parameters $(A,B,C)$ and error scenarios $(a,b,c)$. We found that they maintained their values regardless of the chosen set of $(A-C, a-c)$. We also examined the dependence of the polarization sets of table 1 not only on the different sets of $(A-C, a-c)$, but also on the mass, $M^\alpha$, of the extra gauge boson. We allowed $M^\alpha$ to vary in the range 1.2 TeV $\leq M^\alpha \leq$ 20 TeV for each set of $(A-C, a-c)$ and recalculated the beam polarization sets, $(\lambda_+, \lambda_-)^\alpha_{i,Max}$, $i=1,2,3$, and $(\lambda_+, \lambda_-)^\alpha_{max}$. Our choice of a lower limit of $M^\alpha = 1.2$ TeV is consistent with the present lowest lower search bound on our particles, $M^{LR} \underset{\sim}{>} 1.2$ TeV [10]. We found that the polarization sets of table 1 have also maintained their values independent of the chosen set of $(A-C, a-c)$ and the selected range of $M^\alpha$.

We calculated the 95% C.L. discovery limits, for the extra gauge boson $Z_\alpha$, $\alpha = \chi, \psi, \eta, LR, I'$, using the sets $p = U, P, R, M$ of beam polarizations. In calculating total widths we used assumptions identical to those employed in [11]. In our calculations we considered all different sets, $(A-C, a-c)$, of the chosen colliding beam parameters and error scenarios. The results are represented in Fig. 3 which shows that the highest values for the discovery limits are those obtained using the sets, $(\lambda_+, \lambda_-)^\alpha_{i,Max}$, $i=1,2,3$, as expected. In order to examine the obtained discovery limits we will use them in calculating certain percentage differences. Let $(M^\alpha_d)^{X,x}_{P_1}$ $\left((M^\alpha_d)^{Y,y}_{P_2}\right)$ represents the 95% gauge boson discovery limit obtained for $Z_\alpha$ using set $P_1(P_2)$ of beam polarization and set $(X,x)$ $((Y,y))$ of colliding beam parameters $X(Y)$ and error scenario $x(y)$. We will consider calculating special cases of the percentage difference: $\delta(M^\alpha_d)^{X,x;Y,y}_{P_1,P_2} \equiv \left\{ \left[ (M^\alpha_d)^{X,x}_{P_1} - (M^\alpha_d)^{Y,y}_{P_2} \right] \Big/ (M^\alpha_d)^{Y,y}_{P_2} \right\} \times 100$. Figure 3 shows that, the discovery limits obtained using the sets $(\lambda_+, \lambda_-)^\alpha_{i,Max}$, $i=1,2,3$ differ slightly from those obtained using the single set $(\lambda_+,\lambda_-)^\alpha_{\max}$. In table 2 we present for each model the maximum value of the percentage difference, $\delta(M_d^\alpha)^{X,x;X,x}_{M,\max}$, and the corresponding set, $(X,x)$. The table shows that these maxima have a highest value of no more than $\approx 4.3\%$ which is the case for model $\eta$ and set $(X,x)=(B,c)$. This result suggests -at least for the models under consideration here- the use of the appropriate single collider nominal set ($P$ or $R$) in place of the sets $(\lambda_+,\lambda_-)^\alpha_{i,Max}$, $i=1,2,3$. In fact, it appears impractical to change the beam polarization each time we measure a different observable as the case with the sets $(\lambda_+,\lambda_-)^\alpha_{i,Max}$, $i=1,2,3$. We included in Fig. 3 the discovery limits obtained using set $U$ for comparison. One notices that, for most models, the use of the sets $(\lambda_+,\lambda_-)^\alpha_{\max}$ has greatly enhanced the discovery limits as compared with those limits obtained using set $U$. In fact our results show that the quantity $\delta(M_d^\alpha)^{X,x;X,x}_{\max,U}$ attained a value as high as 47%, which is the case for model $I'$ and set $(X,x)=(A,a)$. The figure also illustrates the role of the inclusion of the systematic errors in reducing the discovery limits. The results obtained using error scenario $a$ put upper bounds on the gauge boson discovery limits. For example, we considered set $B$ and calculated for the models the magnitudes of the percentage reductions in the limits, $\left|\delta(M_d^\alpha)^{B,c;B,a}_{\max,\max}\right|$, due to the inclusion of error scenario $c$. The results are presented in table 2 which show that the percentage reductions ranged from 22% for model $\psi$ to 39% for model $\eta$. Thus, including even a small systematic error reduces the limits substantially. Systematic errors will therefore have to be kept under control. We also examined the dependence of the discovery limits on the collision energy, $\sqrt{s}$, and luminosity, $L$, by calculating the percentage difference, $\delta(M_d^\alpha)^{X,a;Y,a}_{\max,\max}$, for different $(X,Y)$ sets. For example, we found for model $\chi$ that the percentage difference, $\delta(M_d^\alpha)^{X,a;Y,a}_{\max,\max}$, have the following values: $\approx 41\%$ for set $(X,Y)=(C,B)$, $\approx 68\%$ for set $(X,Y)=(C,A)$ and $\approx 19\%$ for set $(X,Y)=(B,A)$. Similar percentages are found for the other models. We compared these percentages with the quantities, $\{[(Ls)_X/(Ls)_Y]^{1/4}-1\}\times 100$, where, for example, $(Ls)_X$ stands for the value of $Ls$ calculated for set $X$ of colliding beam parameters. The comparison showed that our results are consistent with the approximate $(Ls)^{1/4}$ scaling law [12]. The calculations of the gauge boson discovery limits of Fig. 3 are intended to illustrate the role of the proper selection of beam polarization in enhancing the limits. This is clearly demonstrated by the trends of the figure. We note that the calculations did not include effects resulting from the radiative return to $Z_{SM}$ [13]. The inclusion, however, should not affect the trends of Fig. 3. Due to initial state radiation and machine beam-strahlung, the $\sqrt{s}$ spectrum for the process $e^+e^- \to \mu^+\mu^-$ is expected to have a second peak at $\sqrt{s}=M^{SM}$ due to radiative return to $Z_{SM}$ resonance. This is expected to affect the predicted signals from new physics and should therefore be eliminated by appropriate cuts on the energy and angles of the outgoing muons.

In conclusion, an extra neutral gauge boson, $Z_\alpha$, belonging to model $\alpha$, can show up indirectly in $e^+e^-$ collisions through off-resonance deviations of various observables from the corresponding SM values. If the mass of the extra gauge boson is sufficiently large, the expected deviations could be too small to be dealt with. We considered leptonic observables and studied the dependence of the deviations on the polarizations of the positron and electron beams. We showed that these deviations can gain considerable amplification if the polarizations, $\lambda_+$, of the positron beam and $\lambda_-$, of the electron beam, are properly chosen. We considered, as a case study, the neutral gauge bosons $Z_{LR}$ occurring in left-right symmetric models and $Z_\alpha$, $\alpha=\chi,\psi,\eta,I'$, occurring in grand unified theories based on $E_6$ and $SO(10)$ groups (including superstring models). The leptonic observables considered here are the total cross section, $O_1^\alpha \equiv \sigma^\alpha(e^+e^- \to \mu^+\mu^-)$, the forward-backward asymmetry, $O_2^\alpha \equiv A_{FB}^\alpha(e^+e^- \to \mu^+\mu^-)$, and the left-right asymmetry, $O_3^\alpha \equiv A_{LR}^\alpha(e^+e^- \to \mu^+\mu^-)$. Using an algebraic approach, we determined, for a given model $\alpha$, sets of beam polarizations, $(\lambda_+, \lambda_-)_{i,Max}^\alpha$, $i = 1,2,3$, each maximizes the corresponding quantity, $|\Delta O_i^\alpha / \delta O_i^{SM}|$, $i = 1,2,3$, where $\Delta O_i^\alpha$ represents the deviation of $O_i^\alpha$ from $O_i^{SM}$ and $\delta O_i^{SM}$ is the experimental error associated with $O_i^{SM}$. We then showed that when these sets of beam polarizations are used in measuring the corresponding observables, they produced maximal discovery limits, $(M_d^\alpha)_M^{X,y}$, for $Z_\alpha$ where here $X$ and $y$ refer to the employed set of collider beam parameters, $X = (\sqrt{s}, L)$, and error scenario $y$. We similarly determined for each model $\alpha$, a single set of beam polarization, $(\lambda_+, \lambda_-)_{max}^\alpha$, so that if this set is employed in measuring all considered observables, it produces the highest gauge boson discovery limits, $(M_d^\alpha)_{max}^{X,y}$, allowed by a single set of beam polarization. Two interesting results are observed for the considered models. First, each of the sets $(\lambda_+, \lambda_-)_{i,Max}^\alpha$, $i = 1,2,3$, and $(\lambda_+, \lambda_-)_{max}^\alpha$ is found to be one of the beam nominal polarization sets, $P \equiv (|\lambda_+|_{max}, -|\lambda_-|_{max})$ or $R \equiv (-|\lambda_+|_{max}, |\lambda_-|_{max})$. Here $|\lambda_+|_{max}$ and $|\lambda_-|_{max}$ refer to the maximum values of $|\lambda_+|$ and $|\lambda_-|$, respectively. As for $(\lambda_+, \lambda_-)_{max}^\alpha$, we found it equals $P$ for models $\chi$ and $I'$ and equals $R$ for models $\psi$, $\eta$ and $LR$. One could argue that this result is expected, that is, maximum polarizations should be expected to produce highest discovery limits. We, however, stress that this result is obtained for the models studied in the present work and that it should not necessarily be generalized to other models. Moreover, our method enabled us to determine which polarization set, $P$ or $R$, is appropriate for each model. Second, for a given model $\alpha$, the value of the discovery limit, $(M_d^\alpha)_M^{X,y}$, is found to be slightly larger than $(M_d^\alpha)_{\max}^{X,y}$ suggesting the use of the single set $(\lambda_+, \lambda_-)_{\max}^\alpha$ in place of the sets $(\lambda_+, \lambda_-)_{i,Max}^\alpha$, $i = 1,2,3$

.

# Figure Captions

Fig. 1: The quantity $F'^{\alpha}_3 / F'^{\alpha}_1$ calculated for model $\alpha$, where $\alpha = \chi, \psi, \eta, LR, I'$, as a function of the gauge boson mass $M^{\alpha}$. The calculations are made at $\sqrt{s} = 0.5$ TeV and $L = 0.5$ ab$^{-1}$.

Fig. 2: The stack columns for the quantity $(\chi^2)^{\alpha}_p$ calculated for model $\alpha$, where $\alpha = \chi, \psi, \eta, LR, I'$, as a function of the set of beam polarization, $p$, where $p = U, P, R, M$. We take $\sqrt{s} = 0.5$ TeV, $L = 0.5$ ab$^{-1}$ and $M^{\alpha} = 2$ TeV. The polarization sets that correspond to $(\lambda_+, \lambda_-)^{\alpha}_{max}$, are underlined.

Fig. 3: 95% C.L. discovery limits for extra gauge boson, $Z_{\alpha}$, in $e^+e^-$ collisions based on the leptonic observables, $\sigma^{\alpha}(\mu)$, $A^{\alpha}_{FB}(\mu)$ and $A^{\alpha}_{LR}(\mu)$. The calculations are made for models $\alpha$, $\alpha = \chi, \psi, \eta, LR, I'$, sets $U, P, R, M$ of beam polarizations, sets $A, B, C$ of collider beam parameters (orderd respectively from top to bottom for each set of beam polarization), and error scenarios: $a$ (open bars), $b$ (shaded bars) and $c$ (solid bars). The polarization sets that correspond to $(\lambda_+, \lambda_-)^{\alpha}_{max}$, are underlined.

# Table Captions

Table 1: The sets of beam polarizations, $(\lambda_+, \lambda_-)^\alpha_{i,Max}$, $i = 1,2,3$ and $(\lambda_+, \lambda_-)^\alpha_{max}$, for model $\alpha$ where $\alpha = \chi, \psi, \eta, LR, I'$.

Table 2: Maximum value of $\delta(M_d^\alpha)^{X,x;X,x}_{M,\max}$ with the corresponding set $(X, x)$, calculated for all models. Also listed are the magnitudes of the percentage reductions in the discovery limits, $\left|\delta(M_d^\alpha)^{B,c;B,a}_{\max,\max}\right|$, due to the inclusion of error scenario $c$, when set $B$ of collider beam parameters is employed.

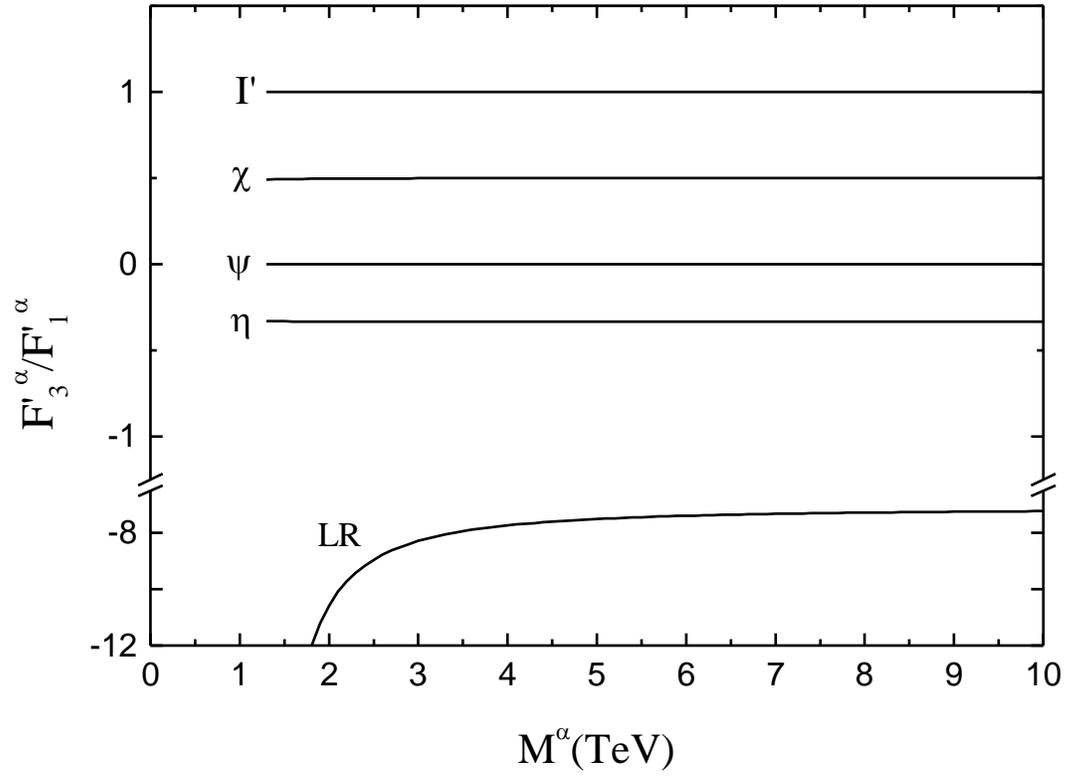

Fig. 1

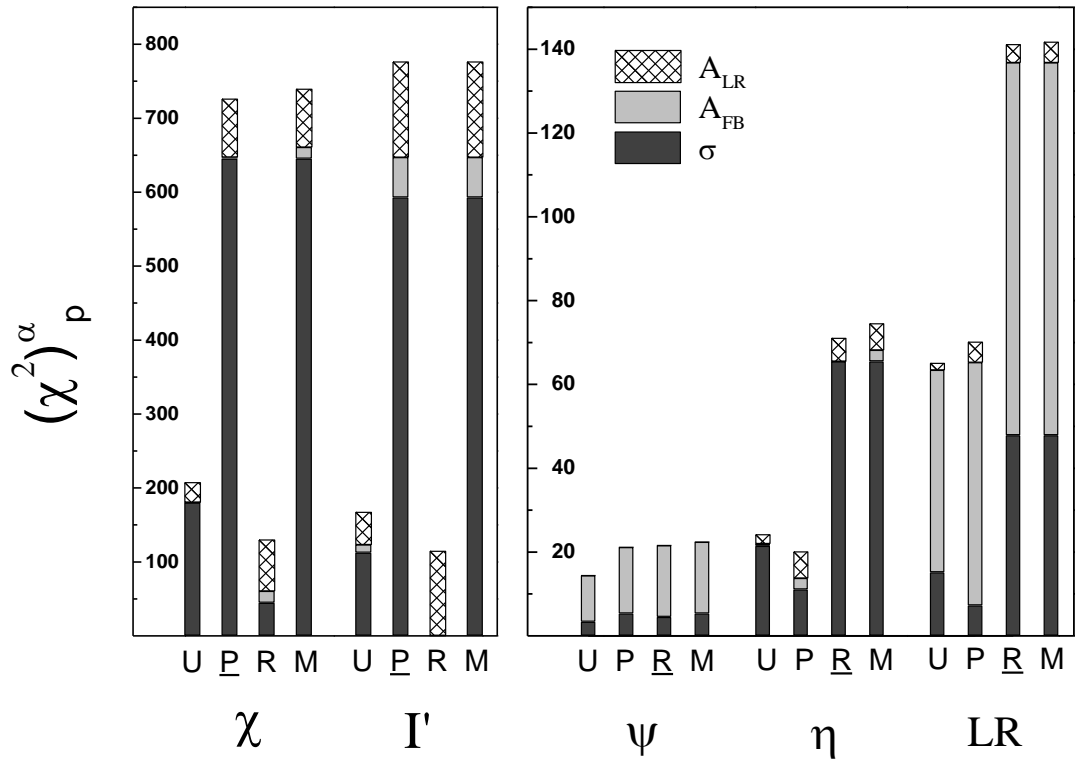

Fig. 2

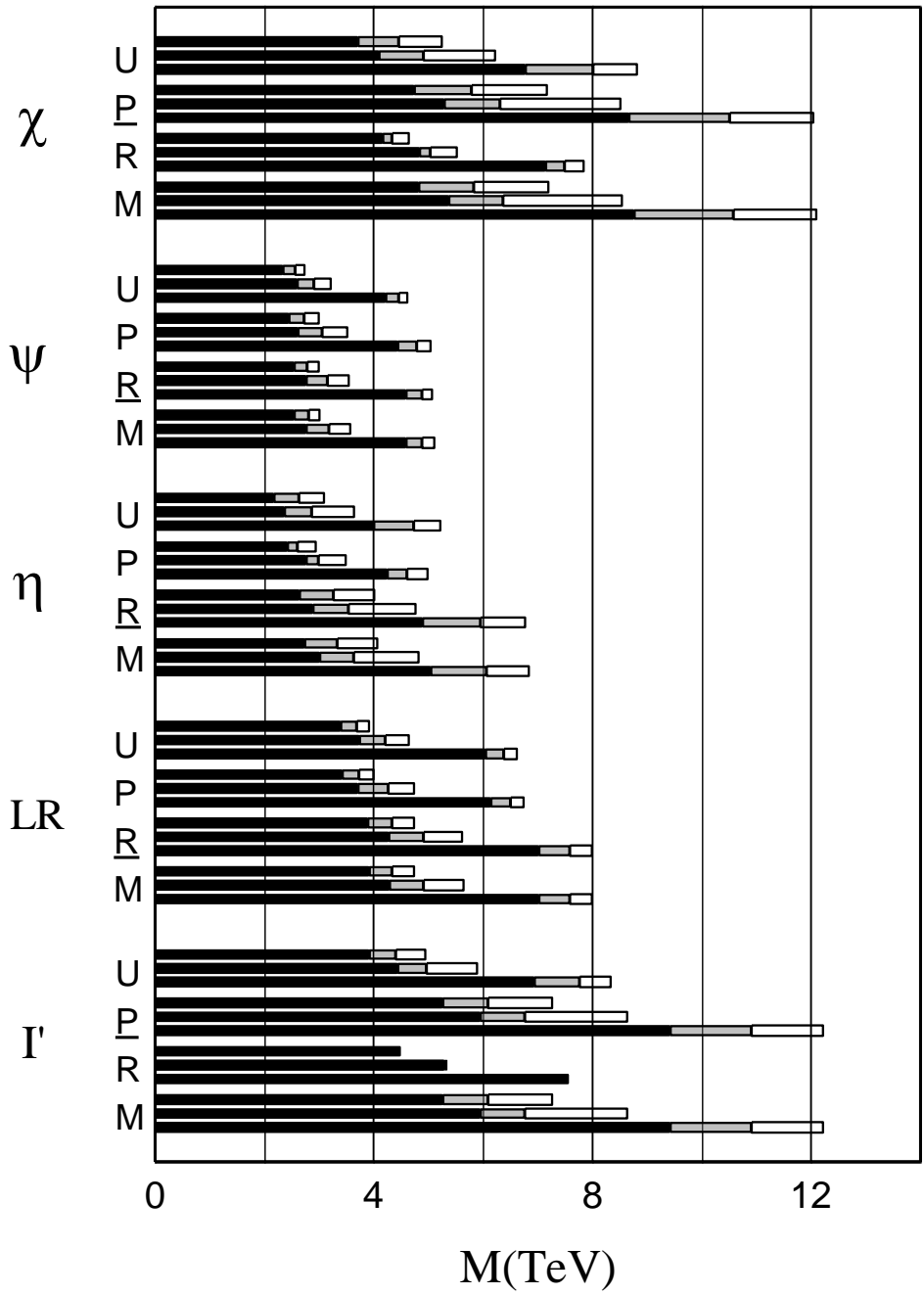

Fig. 3

| Beam Polarization | Model | | | | |
|---|---|---|---|---|---|
| | $\chi$ | $\psi$ | $\eta$ | $LR$ | $I'$ |
| $(\lambda_+, \lambda_-)^{\alpha}_{1,Max}$ | P | P | R | R | P |
| $(\lambda_+, \lambda_-)^{\alpha}_{2,Max}$ | R | R | P | R | P |
| $(\lambda_+, \lambda_-)^{\alpha}_{3,Max}$ | P | P | P | P | P |
| $(\lambda_+, \lambda_-)^{\alpha}_{\max}$ | P | R | R | R | P |

Table 1

|  | Model | | | | |
|---|---|---|---|---|---|
|  | $\chi$ | $\psi$ | $\eta$ | $LR$ | $I'$ |
| $\delta(M_d^\alpha)_{M,\max}^{X,x;X,x}$ | 1.6% | 1% | 4.3% | 0.6% | 0% |
| $(X,x)$ | $(A,c)$ | $(C,a)$ | $(B,c)$ | $(A,c)$ | $(A-C, a-c)$ |
| $\left|\delta(M_d^\alpha)_{\max,\max}^{B,c;B,a}\right|$ | 38% | 22% | 39% | 24% | 31% |

Table 2